\documentclass[aps,prl,twocolumn,citeautoscript,superscriptaddress]{revtex4-2}

\usepackage{graphics}
\usepackage{bigints}
\usepackage{graphicx}
\usepackage{mathrsfs}
\usepackage{amsmath,amssymb}
\usepackage{bm}
\usepackage{color}
\usepackage{float}
\usepackage{lineno}

\makeatletter
\let\frontmatter@footnote@produce\frontmatter@footnote@produce@title
\makeatother



\begin{document}
\title
{Shear Unfreezing Explains Yielding, Plasticity and Neck Initiation of Glassy Polymers}

\author{Peihan Lyu}
\affiliation
{School of Physics, Beihang University, Beijing 100191, China}

\author{Zhaoyu Ding}
\affiliation
{School of Physics, Beihang University, Beijing 100191, China}

\author{Masao Doi}
\affiliation
{School of Physics, Beihang University, Beijing 100191, China}
\affiliation
{Wenzhou Institute, University of Chinese Academy of Sciences, Wenzhou 325000, China}

\author{Xingkun Man}
\email{manxk@buaa.edu.cn}
\affiliation
{School of Physics, Beihang University, Beijing 100191, China}
\affiliation
{Peng Huanwu Collaborative Center for Research and Education, Beihang University, Beijing 100191, China}

\begin{abstract}
		
Yielding, plasticity, and necking are central to the mechanical performance of materials, yet a concise unified physical picture of how these nonlinear responses arise remains lacking. We develop a minimal theory for glassy polymers based on a classical volume-dependent relaxation time following the Doolittle equation, and derive the constitutive relation using the Onsager variational principle. Surprisingly, this simple theory explains yielding, plasticity, and neck initiation under constant strain rate loading via a shear unfreezing mechanism: as the sample is stretched, volume-increasing activated molecular mobility drives shear deformation from an initially frozen state to an unfrozen state. The theory yields an analytical expression for the yielding stress as a function of strain rate and temperature. It also predicts a phase diagram for necking initiation in the same parameter space, providing a mechanism beyond the classical Considère criterion. Our results establish a unified framework for nonlinear tensile behavior in glassy materials.
		
\end{abstract}
	
\maketitle

Nonlinear mechanical responses are ubiquitous in soft and hard materials and are central to materials design and engineering applications~\cite{Sharma2025NPductile,Wang2024NatTough,Cohen2022prlSpider,Diana2019AMLtough}. Glassy polymers provide a canonical example~\cite{Manning2025NatRevPhys,Ediger2024Switch,MFL2023PRL,Fielding2015PRLneck}: under tensile loading, they exhibit a rich sequence of behaviors, including an initial elastic regime, a stress peak at yielding, strain softening, and subsequent strain hardening accompanied by neck formation and propagation. Experiments have revealed that, throughout these processes, the relaxation time of glassy polymer varies by more than three orders of magnitude~\cite{Sergio2022DS_exp,Ediger2009Science}, and that the response depends strongly on temperature and strain rate~\cite{Cheng2013PRLyield,Govaert2012_compress_rate_temperature}.

To rationalize these observations, various theoretical models have been developed. Temperature dependence is typically described by the free-volume concepts~\cite{White2016FreeVolume,Doolittle1951FreeSpace} or the Williams-Landel-Ferry equation~\cite{Wang2018PRLPolymerGlasses,RubinsteinBook_TTSP}, while strain-rate effects are captured by the Eyring~\cite{EyringModel} and Argon~\cite{ArgonModel} models through activated processes. Viscoplastic deformation in amorphous solids is captured by the shear transformation zone concept~\cite{ManningPRL2021ShearTransZone}. History dependence and memory effects are incorporated via internal state variables~\cite{FLC2012prl,ChenSchweizer2011ma}. For necking, the classical Considère criterion associates the onset of instability with the peak in engineering stress, and finite-element simulations
based on constitutive models such as BPA~\cite{BPA1995WuGiessen,BPA1988}, OGR~\cite{OGR1995GlassRubber}, and EGP~\cite{EGP2000JEMT} have been used to study neck propagation. Stability analyses further show that the Considère criterion can fail in soft and viscoelastic materials~\cite{Fielding2017Neck,Fielding2011PRLneck,OGR2010Necking}. Despite these advances, a unified framework that explains yielding, plasticity, and necking from a concise physical mechanism remains lacking.

In this Letter, we propose a minimal continuum model that explains these nonlinear tensile behaviors, based on a model system of glassy polymers. The material is treated as a compressible Kelvin–Voigt viscoelastic solid, with a relaxation time $\tau$ that strongly depends on the sample volume $V$ following the Doolittle equation. This single and classical assumption gives rise to a shear unfreezing mechanism: as volume increases, molecular mobility is activated, driving a transition of shear deformation from a frozen to an unfrozen state. We show that this mechanism naturally and simultaneously explains yielding, plasticity, and the onset of necking.

\begin{figure}[bp]
	\begin{center}
    \includegraphics[width=0.48\textwidth,draft=false]{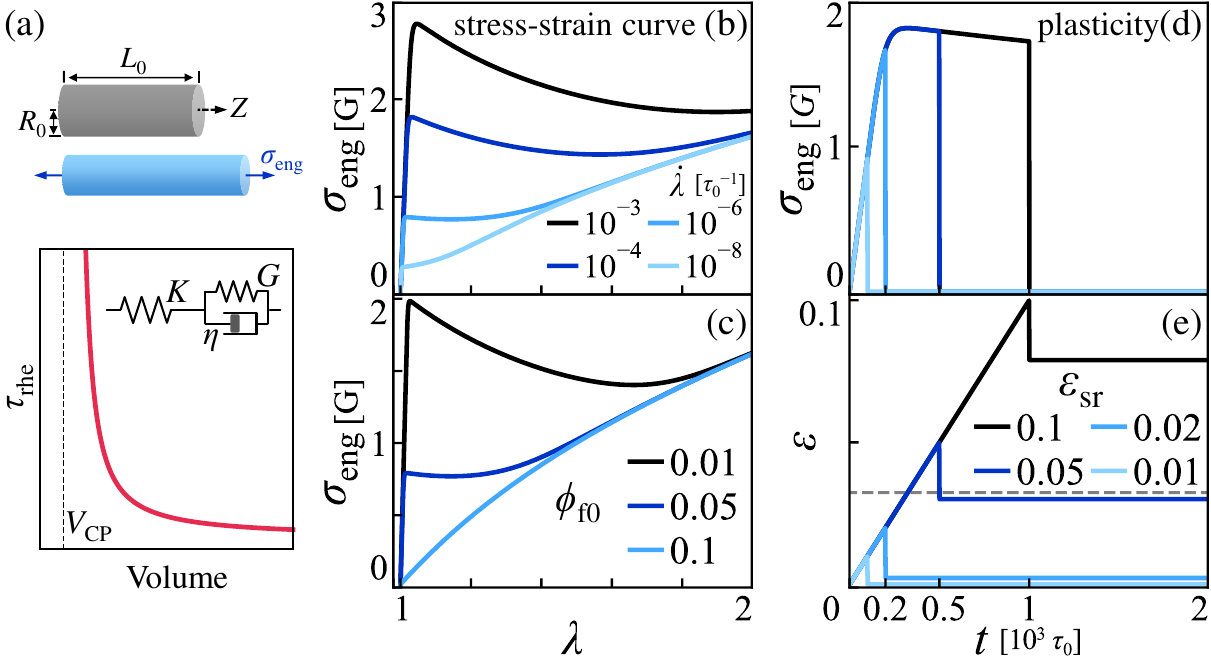}
    \caption{(a) Top: tensile-test geometry. Bottom: constitutive model, where $K$ is the bulk modulus, $G$ is the shear modulus and $\eta$ is the internal viscosity. 
The relaxation time $\tau=\eta/G$, follows Doolittle's equation: it diverges as the local volume approaches close-packing $V_{\rm CP}$ and decreases with increasing volume.
(b, c) Stress-strain curves from constant-strain-rate (CSR) tests: (b) varying strain rate $\dot{\lambda}$ at fixed $\phi_{\rm f0}=0.05$; (c) varying initial free-volume fraction $\phi_{\rm f0}$ at fixed $\dot{\lambda}=10^{-6}\tau_0^{-1}$. 
(d, e) Plasticity: the sample is stretched at constant strain rate $\dot{\lambda}=10^{-4}\tau_0^{-1}$ and $\phi_{\rm f0}=0.05$ to a prescribed strain $\varepsilon_{\rm sr}$, then released under force-free conditions.  
Time evolution of (d) stress and (e) strain for different $\varepsilon_{\rm sr}$. Model parameters: $K/G = 10$.
}
\label{fig1}
\end{center}
\end{figure}

We consider a cylindrical glassy polymer sample of initial length $L_0$, radius $R_0$, and volume $V_0$ in a force-free reference state [Fig.~1(a)]. Under uniaxial tension with homogeneous deformation, a material point at $(R, \Theta, Z)$ moves to $(\sqrt{J/\lambda} R,\, \Theta,\, \lambda Z)$, giving the deformation gradient $\bm{F} = \mathrm{diag}[\sqrt{J/\lambda}, \sqrt{J/\lambda}, \lambda]$. The volume expansion is $J = \mathrm{det}\,\bm{F}$, and the shear (isochoric) deformation part is $\bm{F}_s = J^{-1/3}\bm{F}$. Under constant strain-rate (CSR) loading, $\lambda(t)$ is prescribed, so the deformation is fully characterized by $J(t)$, whose evolution is derived using the Onsager variational principle~\cite{ding2026GeoSwell,lyu2024CaseII}. Then, the engineering stress $\sigma_{\rm eng}$ follows from $J(t)$ and $\lambda(t)$.

We model the material as a compressible Kelvin-Voigt viscoelastic solid ($J \neq 1$), rather than the conventional incompressible assumption, to capture the strong effects of volume change on the mechanical response of glassy polymers. The free energy is
\begin{equation}
A = \int_{V_0} \left[ \frac{1}{2}K(J-1)^2 + \frac{1}{2}G(\bm{F}_s : \bm{F}_s - 3) \right] \mathrm{d}\bm{r} - \sigma_{\rm eng}\lambda V_0,
\end{equation}
where $K$ and $G$ are the bulk and shear moduli, respectively. Dissipation is dominated by shear deformation and takes the form
\begin{equation}
\frac{\Phi}{V_0} = \frac{3}{2}\eta J \left( \frac{\dot{\lambda}}{\lambda} - \frac{\dot{J}}{3J} \right)^2,
\end{equation}
where the viscosity $\eta$ is related to the relaxation time $\tau$ by $\eta = G\tau$. Volumetric dissipation is neglected, as volume changes are treated as purely elastic. 

The model is based on a single classical assumption: a volume-dependent relaxation time $\tau$ [Fig.~1(a)]. Increasing volume enhances molecular mobility, sharply reducing $\tau$ and thus accelerating structural relaxation. Following the Doolittle equation, we write
\begin{equation}
\tau(J) = \tau_0 \exp\left[ \frac{1-\phi_{\rm f0}}{J - (1-\phi_{\rm f0})} \right],
\end{equation}
where $\tau_0$ sets the time scale and $\phi_{\rm f0} = (V_0 - V_{\rm CP})/V_0$ is the initial free-volume fraction, with $V_{\rm CP}$ the close-packed volume. $\phi_{\rm f0}$ serves as an effective temperature parameter, with larger $\phi_{\rm f0}$ corresponding to higher temperatures.

Applying Onsager variational principle, we minimize the Rayleighian $\mathcal{R} = \dot{A} + \Phi$ with respect to $\dot{J}$, yielding the evolution equation of volume expansion,
\begin{equation}\label{eqn:detJ2}
\frac{\dot{J}}{J} = \frac{3\dot{\lambda}}{\lambda} - \frac{1}{\tau} \left[ \frac{3K}{G}(J-1) - J^{-2/3}\!\left(\frac{\lambda^2}{J} - \frac{1}{\lambda}\right) \right].
\end{equation}
The engineering stress is given by
\begin{equation}\label{eqn:sigma_eng}
\sigma_{\rm eng} = \frac{J}{\lambda}\left[ 3K(J-1) \right].
\end{equation}
For a prescribed tensile history $\lambda(t)$, Eq.~\eqref{eqn:detJ2} is solved numerically for $J(t)$, and $\sigma_{\rm eng}$ follows from Eq.~\eqref{eqn:sigma_eng}. For all calculations, stress is scaled by $G$, time by $\tau_0$, and $K/G = 10$ (Poisson's ratio $\nu \approx 0.45$).

Figures~\ref{fig1}(b) and (c) show stress–strain curves under CSR loading, reproducing the characteristic nonlinear response of glassy polymers. At high strain rates $\dot{\lambda}=10^{-3}, 10^{-4}$, and $10^{-6} \, \tau_0^{-1}$, five regimes emerge: (i) linear elasticity, (ii) a yield peak, (iii) strain softening, (iv) plastic flow at nearly constant stress, and (v) strain hardening. At low strain rate $\dot{\lambda}=10^{-8} \,\tau_0^{-1}$, the yield peak and plastic flow vanish, giving a monotonic stress–strain curve. As $\dot{\lambda}$ decreases or $\phi_{\rm f0}$ (temperature) increases, the yield peak decreases and eventually disappears, consistent with experiments~\cite{Ziliang2018ma}.

Figures~\ref{fig1}(d) and (e) show stress-release tests. The sample is stretched at $\dot{\lambda}=10^{-4} \, \tau_0^{-1}$ and $\phi_{\rm f0}=0.05$ to a prescribed strain $\varepsilon_{\rm sr}$, after which the stress is set to $0$. Upon unloading, the extensional strain $\varepsilon$ relaxes to a finite residual value. Such results demonstrate that the model captures the irreversible plastic deformation on observable timescales.

\begin{figure}[bp]
	\begin{center}
		\includegraphics[width=0.47\textwidth,draft=false]{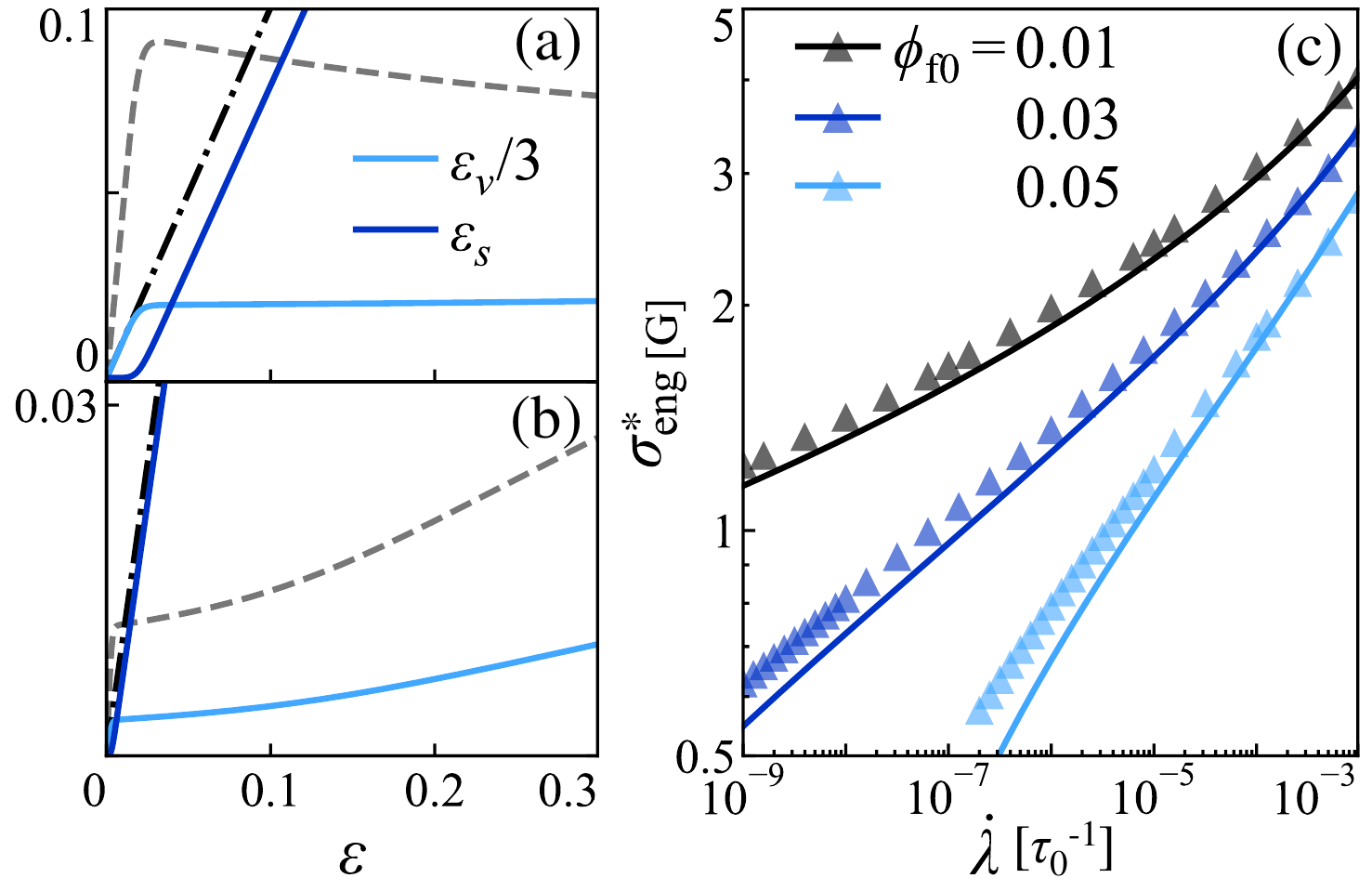}
		\caption{Evolution of volumetric strain $\varepsilon_{\rm v}$ and shear strain $\varepsilon_{\rm s}$ during CSR tests for (a) fast and (b) slow strain rates, $\dot{\lambda}=10^{-4}\tau_0^{-1}$ and $10^{-8}\tau_0^{-1}$, with $\phi_{\rm f0}=0.05$. Gray dashed lines show $\sigma_{\rm eng}$, while black dash-dotted lines show the extensional strain $\varepsilon$. (c) Yield stress: comparison between the full model (triangles) and the approximation $\sigma_{\rm eng}^*=3K\varepsilon_{\rm v}^*$ computed with $C=\ln 0.1$.}
		\label{fig2}
	\end{center}
\end{figure} 

\textit{Yielding.---}
We first analyze yielding. For clarity, we define the extensional strain as $\varepsilon=\lambda-1$,  and decompose it into volumetric and shear (isochoric) components, $\varepsilon_{\rm v}=J-1$ and $\varepsilon_{\rm s}=\lambda J^{-1/3}-1$, respectively. The engineering stress (Eq.~\eqref{eqn:sigma_eng}) can then be rewritten as
\begin{equation}\label{eqn:small_sigma_eng}
\sigma_{\rm eng}=\frac{3K\varepsilon_{\rm v}(1+\varepsilon_{\rm v})^{\frac{2}{3}}}{1+\varepsilon_{\rm s}}.
\end{equation}
This expression shows that the stress evolution is governed by the coupled dynamics of $\varepsilon_{\rm v}$ and $\varepsilon_{\rm s}$.

Figure~\ref{fig2}(a) shows the strain evolution at a high strain rate ($\dot{\lambda}=10^{-4}\tau_0^{-1}$), exhibiting a typical yielding response (gray dashed line). Before yielding, $\varepsilon_{\rm v}$ increases rapidly and closely follows the extensional strain $\varepsilon$ (black dash-dotted line), while shear deformation is suppressed ($\varepsilon_{\rm s}\approx 0$), defining a shear-frozen state. In this regime, Eq.~\eqref{eqn:small_sigma_eng} reduces to $\sigma_{\rm eng}\approx 9K\varepsilon$~\cite{SI}, giving a linear elastic-like response consistent with experiments~\cite{bookchap2016polymglass}.  Beyond yielding, $\varepsilon_{\rm v}$ remains nearly constant, whereas $\varepsilon_{\rm s}$ starts to increase rapidly, defining a shear-unfrozen state. This shear frozen-to-unfrozen transition is driven by the sharp decrease of $\tau$ due to increasing $\varepsilon_{\rm v}$. As $\varepsilon_{\rm s}$ grows while $\varepsilon_{\rm v}$ stays nearly constant, Eq.~\eqref{eqn:small_sigma_eng} predicts a stress decrease, corresponding to stress softening and yielding.

Fig.~\ref{fig2}(b) shows a slow strain-rate case ($\dot{\lambda}=10^{-8}\tau_0$). In contrast to Fig.~\ref{fig2}(a),  no yielding occurs. Here, $\varepsilon_{\rm s}$ exhibits essentially no frozen regime and increases continuously. Meanwhile, $\varepsilon_{\rm v}$ increases gradually rather than remains constant after a rapid initial increase. As a result, the stress keeps increasing and no yielding is observed.

We propose a shear unfreezing mechanism for yielding. The material evolves from a shear-frozen state ($\varepsilon_{\rm s}\approx 0$) to a shear-unfrozen state where $\varepsilon_{\rm s}$ increases rapidly. In the frozen state, the material has high viscosity and low molecular mobility (large $\tau$); deformation is primarily volumetric, leading to a monotonic stress increase. As deformation proceeds, $\tau$ decreases sharply, reducing viscosity and activating molecular mobility, which drives the rapid growth of shear deformation $\varepsilon_{\rm s}$. This transition releases the stress accumulated in the shear-frozen state and gives rise to yielding.

The model also predicts the yielding strain $\varepsilon^*_v$ and yielding stress $\sigma^*_{\rm eng}$. The yielding point often occurs in the small-deformation regime ($\lambda-1\ll1$), leading to $\varepsilon \approx \varepsilon_{\rm v}/3 + \varepsilon_{\rm s}$ and $\dot{\varepsilon}_{\rm s} \approx \dot{\lambda}_{\rm s}/\lambda_{\rm s}$. Equation~\eqref{eqn:detJ2} then reduces to
\begin{equation}\label{eqn:vol_strain_approx}
\dot\varepsilon_{\rm v} \approx 3\left[ \dot\varepsilon - \frac{1}{\tau(\varepsilon_{\rm v})}\left(\frac{3K+G}{3G}\varepsilon_{\rm v}-\varepsilon\right) \right].
\end{equation}
At yielding point, the stress reaches a maximum, $\dot{\sigma}_{\rm eng}=0$, while shear deformation remains negligible, $\varepsilon_{\rm s}\approx 0$. Substituting the two conditions into Eq.~\eqref{eqn:vol_strain_approx}, we have
\begin{equation}\label{eqn:vol_strain_yielding}
\varepsilon_{\rm v}^*=\frac{1}{\ln(K/G)+C-\ln(\dot\varepsilon\tau_0)}-\phi_{\rm f0},
\end{equation}
where $C=\ln\varepsilon_{\rm v}^*$ is treated as a constant and $1-\phi_{\rm f0}$ is replaced by $1$. Eq.~\eqref{eqn:vol_strain_yielding} indicates that $\varepsilon_{\rm v}^*$ changes with inverse of $\ln(\dot\varepsilon\tau_0)$, which explain the weak dependence of $\varepsilon_{\rm v}^*$ on $\dot\varepsilon$.

The yielding stress is then given as $\sigma_{\rm eng}^*\approx 3K\varepsilon_{\rm v}^*$. Figure~\ref{fig2}(c) compares this prediction in Eq.~\eqref{eqn:vol_strain_yielding} with numerical results, showing good agreement. The model further captures the increase of $\sigma_{\rm eng}^*$ with $\dot{\lambda}$ and its decrease with $\phi_{\rm f0}$, consistent with experimental results~\cite{Govaert2012_compress_rate_temperature}.

\begin{figure}[tp]
	\begin{center}
    \includegraphics[width=0.47\textwidth,draft=false]{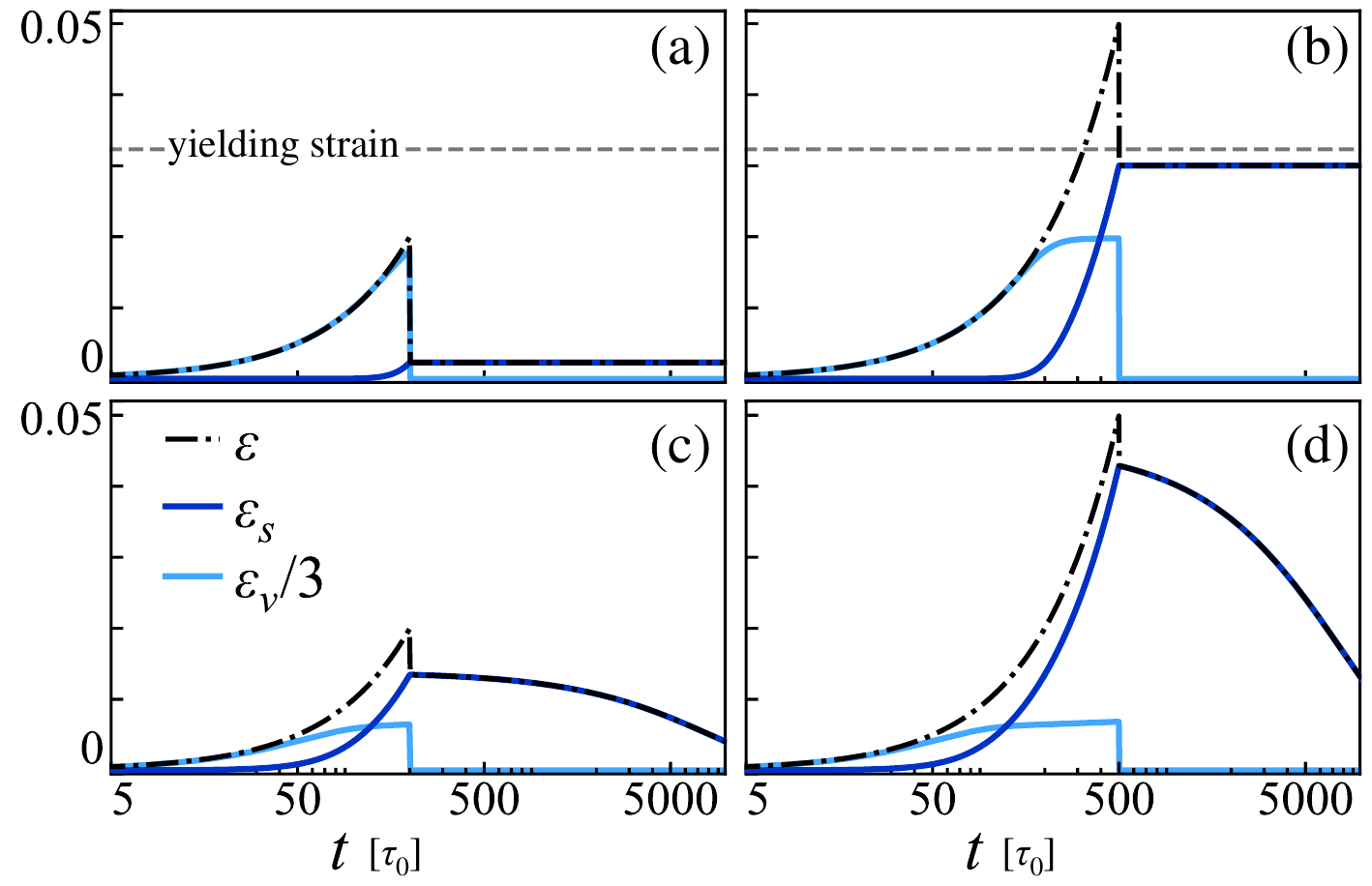}
    \caption{Time evolution of strain components during stress-release tests. Black dash-dotted, dark-blue solid, and light-blue solid curves denote total strain $\varepsilon$, shear strain $\varepsilon_{\rm s}$, and volumetric strain $\varepsilon_{\rm v}$, respectively. Top row: $\phi_{\rm f0}=0.05$ (lower temperature). Stress release at (a) $\varepsilon_{\rm sr}=0.02$ (pre-yielding) and (b) $\varepsilon_{\rm sr}=0.05$ (beyond-yielding). Bottom row: $\phi_{\rm f0}=0.1$ (higher temperature). (c) $\varepsilon_{\rm sr}=0.02$, and (d) $\varepsilon_{\rm sr}=0.05$. Gray-dashed lines indicate the yielding strain for guidance. For all calculations, the strain rate is $\dot{\lambda}=10^{-4}\tau_0^{-1}$.
    }
		\label{fig3}
	\end{center}
\end{figure}

\textit{Plasticity.---}
We turn to analyze plasticity. Figure~\ref{fig3} shows the evolution of $\varepsilon$ and its components $\varepsilon_{\rm s}$ and $\varepsilon_{\rm v}$ during stress-release tests. We consider two temperatures, $\phi_{\rm f0}=0.05$ (lower temperature) and $\phi_{\rm f0}=0.1$ (higher temperature), and two release strains, $\varepsilon_{\rm sr}=0.02$ (before yield) and $\varepsilon_{\rm sr}=0.05$ (beyond yield).

For the low-temperature cases [Figs.~\ref{fig3}(a) and (b)], the behavior is similar for both values of $\varepsilon_{\rm sr}$. Upon stress release, $\varepsilon$ drops suddenly to a finite residual value and then remains nearly constant. The volumetric strain $\varepsilon_{\rm v}$ drops immediately to zero, while the shear strain $\varepsilon_{\rm s}$ stays frozen at the value reached at the moment of release. The residual strain is therefore entirely carried by $\varepsilon_{\rm s}$. 

Interestingly, plastic residual strain appears even before yielding [Fig.~\ref{fig3}(a)], indicating that the material is not purely elastic prior to yielding. This is consistent with recent experiments~\cite{Woodbridge2026PRLbeforeyield}.

For the high-temperature cases [Figs.~\ref{fig3}(c) and (d)], $\varepsilon_{\rm v}$ again drops to $0$ upon stress release. However, both $\varepsilon$ and $\varepsilon_{\rm s}$ now gradually relax to $0$ rather than remaining constant. This indicates that the strain is fully recoverable and the material exhibits no permanent plasticity.

The sudden drop of $\varepsilon_{\rm v}$ upon stress release reflects the purely elastic nature of volumetric deformation, which arises naturally from the model. Setting $\sigma_{\rm eng}=0$ in Eq.~\eqref{eqn:small_sigma_eng} immediately gives $\varepsilon_{\rm v}=0$, independent of its value before release.

In contrast, the evolution of $\varepsilon_{\rm s}$ follows
\begin{equation}
\varepsilon_{\rm s}(t)=\varepsilon^{\rm s}_{\rm sr}\exp\left[-\frac{t-t_{\rm sr}}{\left.\tau\right|_{\varepsilon_{\rm v}=0}}\right],
\end{equation}
where $\varepsilon^{\rm s}_{\rm sr}$ is the shear strain at the release time $t_{\rm sr}$~\cite{SI}. At the instant of release, $\varepsilon_{\rm v}=0$ ($J=1$), then the relaxation time becomes $\left.\tau\right|_{\varepsilon_{\rm v}=0}=\tau_0 \exp[(1-\phi_{\rm f0})/\phi_{\rm f0}]$. For small $\phi_{\rm f0}$ (low temperature, $T<T_g$), $\tau$ is extremely large. This indicates that the system returns to a shear-frozen state again, leading to permanent residual strain. For larger $\phi_{\rm f0}$ (higher temperature), $\tau$ decreases, and $\varepsilon_{\rm s}$ relaxes over time, resulting in strain recovery. This explains the transition from irreversible plasticity below $T_g$ to reversible rubbery behavior as $T$ approaches or exceeds $T_g$.

\textit{Necking.---}
In many tensile tests, homogeneous deformation becomes unstable and strain localization eventually leads to necking. To analyze this, we introduce a small perturbation to the deformation gradient. A material point initially at $(R, \Theta, Z)$ moves to $(\lambda_R(t)R,\, \Theta,\, \lambda(t)Z) + \alpha(\delta r,\, 0,\, \delta z)$ at time $t$, where $\alpha \ll 1$ and $\alpha\delta r$, $\alpha\delta z$ are perturbations in the radial and axial directions, respectively. We adopt a long-wavelength approximation, retaining only perturbations on scales larger than the sample radius, which gives $\delta r = \rho(Z,t)R$ and $\delta z = \zeta(Z,t)R_0$. The sample shape is then fully described by $J(t)$, $\rho(Z,t)$, and $\zeta(Z,t)$. Following standard stability analysis, we write $\rho = k_2 \exp(st)\cos(k_Z Z)$ and $\zeta = k_1 \exp(st)\sin(k_Z Z)$, where $s$ is the growth rate, $k_1$ and $k_2$ are amplitudes, and $k_Z = n\pi/L_0$ is the wave number fixed by the boundary conditions~\cite{SI}. Then, we use the Onsager variational principle to determine the three variables $J$, $\rho$, and $\zeta$, which gives the evolution of the growth rate $s(t)$.

Figure~4 is the evolution of $s(t)$. At low temperatures $\phi_{\rm f0}=0.03$ and $0.05$ or high strain rate $\dot{\lambda}=10^{-4}\tau_0^{-1}$, $s(t)$ remains positive and of order $\dot{\lambda}$ at early times, then rises rapidly to a peak near yielding (star points) and gradually decays. In contrast, at high temperature $\phi_{\rm f0}=0.08$ or low strain rate $\dot{\lambda}=10^{-8}\tau_0^{-1}$, $s(t)$ quickly becomes negative, indicating suppression of perturbations. The peak value of $s$ increases with increasing $\dot{\lambda}$ and decreasing $\phi_{\rm f0}$, showing that faster stretching and lower temperature promote perturbation growth and neck initiation. Here, we do not consider brittle fracture: for glassy polymers at temperatures far below $T_{\rm g}$ and extremely high strain rates, the response transitions to brittle failure rather than necking.

This behavior follows from stress accumulation before and release after shear unfreezing. Before unfreezing, shear strain is effectively frozen and stress accumulates. As shown in Fig.~2(c), the yield stress $\sigma^*_{\rm eng}$ increases with $\dot{\lambda}$ and decreasing $\phi_{\rm f0}$, consistent with the trend of $s(t)$. Thus higher strain rates and lower temperatures produce larger stress accumulation prior to yielding. Once shear unfreezing occurs, the stored stress is released, amplifying perturbations and promoting necking.

\begin{figure}[tp]
	\begin{center}
    \includegraphics[width=0.47\textwidth,draft=false]{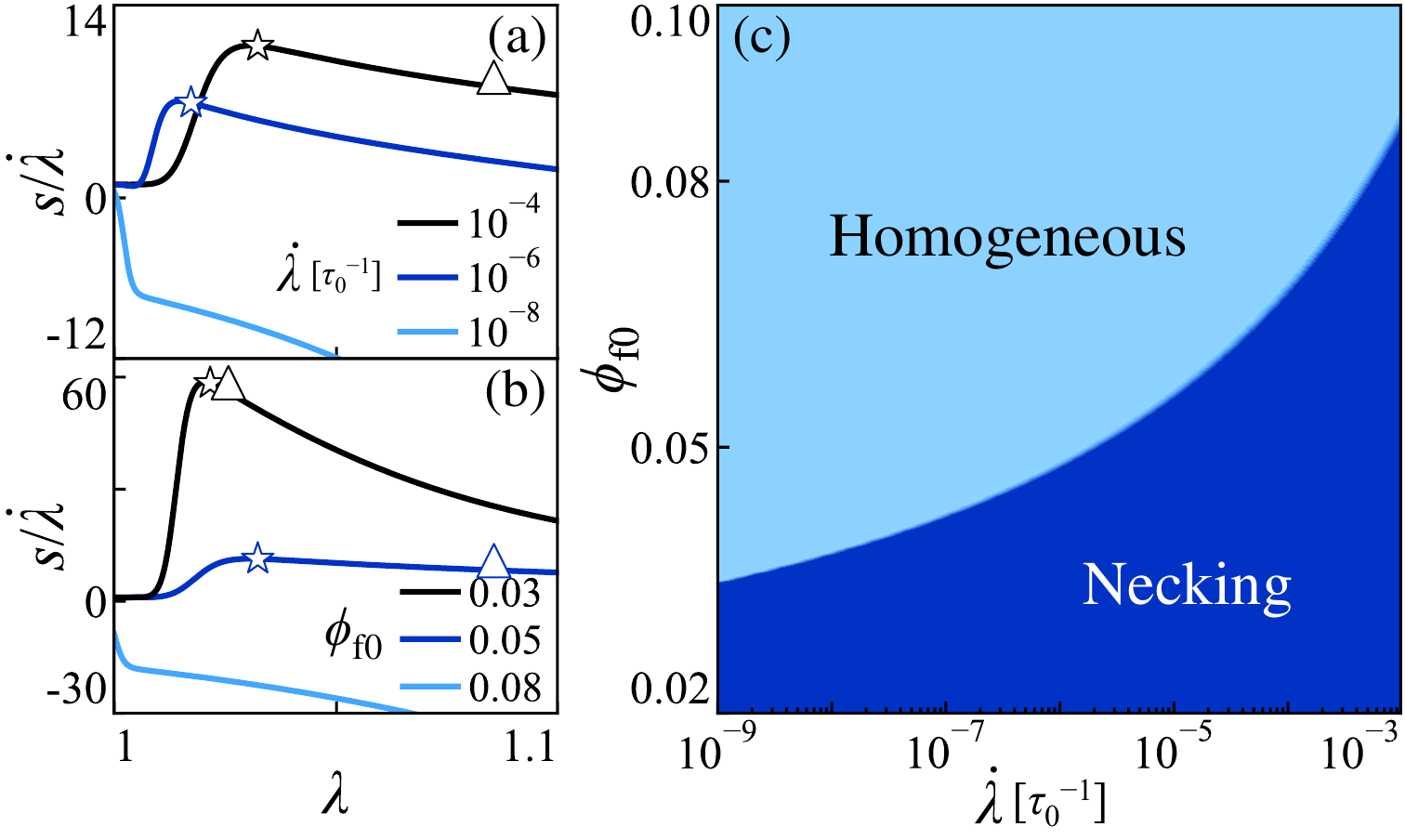}
    \caption{Time evolution of perturbation growth rate $s$ in CSR tests: (a) varying $\dot{\lambda}$ at fixed $\phi_{\rm f0}=0.05$; (b) varying $\phi_{\rm f0}$ at fixed $\dot{\lambda}=10^{-4}\tau_0^{-1}$. (c) Phase diagram in terms of $\dot{\lambda}$ and $\phi_{\rm f0}$, separating the necking phase (dark blue) from the homogeneous deformation phase (light blue), with phase boundary determined by $M_c=2$.
  }
		\label{fig4}
	\end{center}
\end{figure}

However, the instantaneous growth rate $s(t)$ alone is insufficient to determine neck initiation. Although $s$ remains positive at early times, no necking is observed experimentally before yielding point. This is because the accumulated amplification is limited, i.e., it is too weak to produce appreciable strain localization~\cite{LiBuckley2009IJSS,Fielding2016JRheoP2}.

To quantify this effect, we introduce an amplification factor of perturbations over deformation history
\begin{equation}
M(t)=\exp\left(\int_0^t s(t')\,{\rm d}t'\right).
\end{equation}
We further define a critical threshold $M_c$: necking occurs when $M > M_c$, whereas the deformation remains homogeneous if $M < M_c$. Fig.~4(c) shows the phase diagram of neck initiation in the $\dot{\lambda}$ and $\phi_{\rm f0}$ space based on this criterion, referred to as the M-criterion. Necking occurs in the high strain rate and low temperature regime as long as brittle failure does not take place. 

The M-criterion differs from the classical Considère criterion. The latter attributes necking to strain softening near yielding, whereas the M-criterion accounts for the full evolution from stress accumulation in the shear-frozen state to its release upon shear unfreezing, thereby incorporating history-dependent effects. Consequently, the onset of necking predicted by the M-criterion is not restricted to the vicinity of yielding. This is shown by the triangular data points in Figs.~4(a) and (b). 
Necking onset given by M-criterion coincide with the yielding point only in fast strain-rate and low temperature case ($\phi_{\rm f0}=0.03$ in fig.4(b)), but deviate from yielding point as temperature increases.
Notably, yielding does not necessarily lead to necking within the M-criterion. For instance, at $\dot{\lambda}=10^{-6} \, \tau^{-1}_0$ in Fig. 4(a), the response remains homogeneous despite yielding. Similar behavior has also been reported in recent experiments~\cite{Sergio2022DS_exp} and theories on soft glassy polymers~\cite{Fielding2015PRLneck}.

In summary, we develop a minimum theory for the tensile response of glassy polymers that explains yielding, plasticity, and neck initiation. Based on a strongly volume-dependent relaxation time, we propose a unified shear unfreezing mechanism for nonlinear mechanical responses in tensile tests, in which deformation-activated molecular mobility drives a transition of shear deformation from a frozen state to an unfrozen state. The theory gives an analytical expression for the yielding stress as a function of strain rate and temperature, and explains plastic irreversibility via re-entry into the shear-frozen state upon unloading. Combined with perturbation analysis, it predicts that higher strain rates and lower temperatures promote necking initiation via enhanced stress accumulation in the shear-frozen state and its rapid release during shear unfreezing process. The model can be systematically extended to incorporate non-Gaussian chain elasticity and more complex dissipation pathways, enabling quantitative descriptions of neck propagation, strain-rate-dependent stress overshoot~\cite{nanzai1990exp}, compressive responses~\cite{LIU2022PC_compress,dar2014PC_compress}, and multistep loading protocols~\cite{Medvedev2022multistep}.

\acknowledgments{
\textit{Acknowledgment.---}This work was supported in part by the National Natural Science Foundation of China (NSFC) under Grant No. 22473005, No. 21961142020, and No. 21822302, the Fundamental Research Funds for the Central Universities under Grant No. YWF-22-K-101, and the China Scholarship Council (CSC). The authors also acknowledge the support of the High Performance Computing Center of Beihang University.
}


%

\end{document}